\documentclass[%
 reprint,
 amsmath,amssymb,
 aps,
]{revtex4-2}

\usepackage{graphicx}
\usepackage{dcolumn}
\usepackage{bm}
\usepackage[mathlines]{lineno}
\usepackage{braket}
\usepackage[dvipsnames]{xcolor}
\usepackage{amsmath,amssymb,amsfonts} 
\usepackage[utf8]{inputenc}
\usepackage{appendix}
\usepackage{mhchem}
\usepackage{makecell}
\usepackage{float}
\usepackage{bbm}
\usepackage{nicefrac}
\usepackage{wasysym}
\usepackage[normalem]{ulem}

\usepackage{hyperref}
\hypersetup{
	colorlinks=true,
	linkcolor=blue,
	filecolor=blue,      
	urlcolor=blue,
	citecolor=blue,
}

\begin{document}

\preprint{APS/123-QED}

\author{Yixin Zhang}
\affiliation{School of Physics, Peking University, Beijing 100871, China}

\author{H. Huang}
\email[Contact author: ]{huanghq07@pku.edu.cn} 
\affiliation{School of Physics, Peking University, Beijing 100871, China}

\date{\today}

\title{Optical Hall absorption sum rule and spectral compensation in time-reversal-breaking moir\'e and Hofstadter systems}

\date{\today}

\begin{abstract}
Optical spectroscopy provides a powerful, contact-free probe of topological quantum states, yet exact constraints on antisymmetric Hall absorption remain much less well developed than their longitudinal counterparts. Motivated by earlier Hall-conductivity sum rules, we formulate the corresponding first-frequency-moment constraint for the antisymmetric optical conductivity, whose imaginary part governs chirality-dependent absorption. We then demonstrate this sum rule in two classes of time-reversal-breaking topological systems. For a zero-field moiré continuum model hosting topological bands, the moment vanishes exactly, implying that any low-frequency anomalous Hall absorption must be compensated by higher-frequency spectral weight of the opposite sign. For a Hofstadter model under a uniform magnetic field, the same moment takes a universal value fixed by the magnetic flux density, independent of microscopic model details. By linking low- and high-frequency spectral contributions, this optical Hall absorption sum rule provides a rigorous framework for quantifying circular dichroism constraints and diagnosing Landau-level mixing. Our results show how a known Hall spectral constraint acquires new and experimentally relevant content in modern interacting topological materials.
\end{abstract}

\maketitle

\textit{Introduction.}---Optical spectroscopy offers a powerful, contact-free probe of topological quantum matter~\cite{tholeXrayCircularDichroism1992, wuQuantizedFaradayKerr2016, dziomObservationUniversalMagnetoelectric2017, moligniniProbingChernNumber2023, caiSignaturesFractionalQuantum2023, taoValleyCoherentQuantumAnomalous2024, bacProbingBerryCurvature2025}. While the dc Hall response is governed by Berry curvature and band topology~\cite{klitzingNewMethodHighAccuracy1980, jungwirthAnomalousHallEffect2002, onodaTopologicalNatureAnomalous2002, haldaneBerryCurvatureFermi2004}, its finite-frequency counterpart contains additional information  for characterizing orbital magnetization~\cite{tholeXrayCircularDichroism1992, souzaDichroic$f$sumRule2008}, identifying band topology~\cite{wuQuantizedFaradayKerr2016, okadaTerahertzSpectroscopyFaraday2016, moligniniProbingChernNumber2023, onishiFundamentalBoundTopological2023, ghoshProbingQuantumGeometry2024, vermaInstantaneousResponseQuantum2024, yuUniversalWilsonLoop2025, bacProbingBerryCurvature2025}, diagnosing correlated insulators~\cite{uchidaOpticalSpectra$mathrmLa_2mathrmensuremathmathitx$$mathrmSr_mathitx$$mathrmCuO_4$1991,  katsufujiSpectralWeightTransfer1995, rozenbergTransferSpectralWeight1996}, and elucidating exciton physics in topological matter~\cite{caiSignaturesFractionalQuantum2023, taoValleyCoherentQuantumAnomalous2024}. Sum rules provide especially useful constraints by linking integrated optical response to equal-time ground-state properties. The best-known example is the longitudinal f-sum rule, linking the frequency-integrated optical longitudinal conductivity to total carrier density~\cite{reicheUeberZahlDispersionselektronen1925, giulianiQuantumTheoryElectron2005}. More recently, related ideas have been extended toward geometric bounds, where optical weight is linked to quantum metric, Chern number, and other characteristics of topological quantum states~\cite{onishiFundamentalBoundTopological2023, onishiQuantumWeight2024, onishiTopologicalBoundStructure2024, onishiUniversalRelationEnergy2024, vermaInstantaneousResponseQuantum2024, qiuQuantumGeometryProbed2024, souzaOpticalBoundsManyelectron2025, passosElectronicBoundsMagnetic2025}. 

By comparison, exact constraints on Hall-related optical response are less extensively developed. Nevertheless, several important precedents exist. Earlier works established exact constraints for the optical Hall angle, the ac Hall constant, and, notably, the first frequency moment of the Hall conductivity in correlated electron systems near the Mott transition~\cite{drewSumRuleOptical1997, langeMemoryfunctionApproachHall1997, langeMagnetoopticalSumRules1999, souzaDichroic$f$sumRule2008}. These results show that antisymmetric optical response is also subject to rigorous spectral constraints. However, their implications for modern time-reversal-breaking topological systems, especially moir\'e continuum bands at zero external field and Hofstadter bands in a uniform magnetic field, have remained largely unexplored.

Here we revisit this Hall first-moment constraint in the language of optical Hall absorption and circular dichroism. We formulate the sum rule for the antisymmetric optical conductivity $\tilde{\sigma}_{xy}(\omega)$, whose imaginary part controls chirality-resolved absorption, and evaluate it in two distinct topological settings. For a zero-field moir\'e continuum model, we show that the relevant current commutator vanishes exactly, implying that the net first moment of Hall absorption must vanish even when the low-energy anomalous Hall optical response is finite. For a Hofstadter model under a uniform magnetic field, by contrast, the same moment takes a universal value fixed by the magnetic flux density. These results establish a direct link between circular dichroism, spectral-weight compensation across energy scales, and Landau-level mixing, and show how the Hall first-moment sum rule can serve as a diagnostic of the microscopic origin of topological optical response.

\begin{figure}[htbp]
\includegraphics[width=1\linewidth]{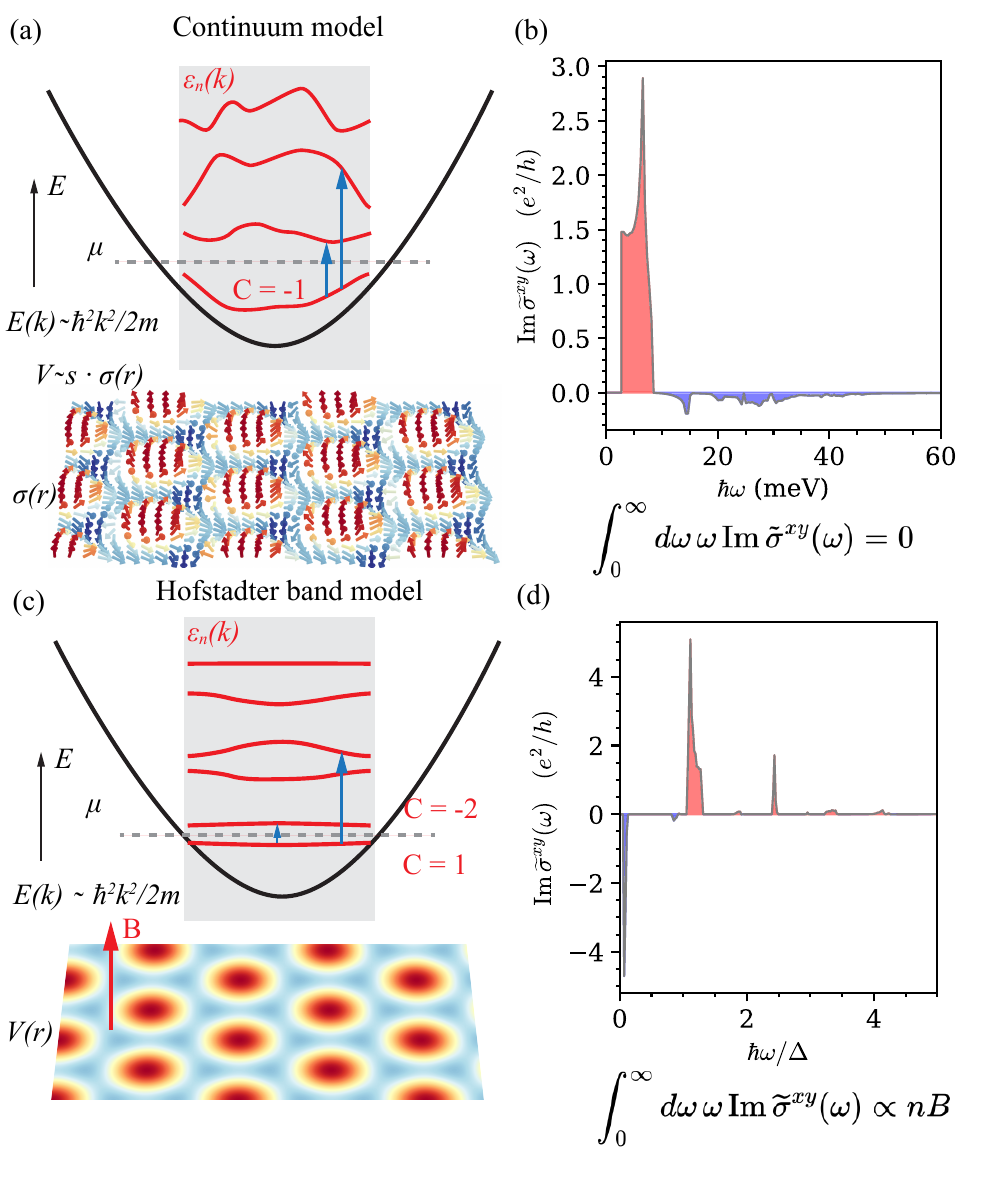}
\caption{\label{fig:1} (a) Continuum model description: electrons move in a skyrmion texture generating a position-dependent Zeeman field. The free-electron parabolic dispersion is folded into the moir\'e Brillouin zone, yielding schematic folded bands $\varepsilon_n(\boldsymbol{k})$ (red). The gray dashed line indicates the chemical potential $\mu$. (b) Schematic optical Hall response for the system in (a). The first frequency moment of the antisymmetric Hall conductivity vanishes identically, $\int_{0}^{\infty} d\omega\, \omega\, \mathrm{Im}\,\tilde{\sigma}_{xy}(\omega)=0$, where $\tilde{\sigma}_{xy}(\omega)\equiv \frac{1}{2}[\sigma_{xy}(\omega)-\sigma_{yx}(\omega)]$. (c) Hofstadter model: electrons subject to a uniform magnetic field $B$ and a spatially periodic potential $V(\boldsymbol{r})$. The resulting Landau levels are broadened and split (schematic $\varepsilon_n(\boldsymbol{k})$ in red). Chemical potential is set to occupy only the lowest subband. (d) Schematic optical Hall response for (c). The first moment satisfies $\int_{0}^{\infty} d\omega\, \omega\, \mathrm{Im}\,\tilde{\sigma}_{xy}(\omega)= \pi q^3 n B /(2 m^2)$. Here, \(\Delta = \hbar |e B| / m\) denotes the clean cyclotron gap.}
\end{figure}

\textit{Sum rule derivation}---We now formulate the first-moment Hall sum rule in a form tailored to optical Hall absorption. The derivation is closely related in spirit to earlier Hall-conductivity sum rules, particularly the first-moment relation obtained by Lange and Kotliar for correlated electron systems~\cite{langeMagnetoopticalSumRules1999}. For completeness, and to establish notation suited to antisymmetric optical response in topological materials, we briefly recast the argument here for the antisymmetrized conductivity
$\tilde{\sigma}_{xy}(\omega) = \frac{1}{2}[\sigma_{xy}(\omega) - \sigma_{yx}(\omega)]$, whose imaginary part directly governs chirality-dependent absorption. We begin with the Kubo-Greenwood formula for the complex conductivity $\sigma_{\mu\nu}(z)$ in the upper half-plane ($\operatorname{Im} z > 0$):
\begin{equation}
\sigma_{\mu\nu}(z) = \frac{i}{V z} \left[ \langle \tau_{\mu\nu} \rangle + \frac{1}{i\hbar} \int_0^\infty dt \, e^{i z t} \langle [\hat{J}_\mu(t), \hat{J}_\nu(0)] \rangle \right].
\end{equation}
Here, $V$ is the volume, and $\langle \cdot \rangle$ denotes the thermodynamic average. For the antisymmetric component $\tilde{\sigma}_{xy}(z)$, the diamagnetic term $\langle \tau_{\mu\nu} \rangle$ (symmetric in $\mu, \nu$) vanishes, yielding:
\begin{equation}
\tilde{\sigma}_{xy}(z) = \frac{1}{\hbar V z} \int_0^\infty dt \, e^{i z t} \langle [\hat{J}_x(t), \hat{J}_y(0)] \rangle.
\end{equation}

Expanding $\hat{J}_x(t) \approx \hat{J}_x(0) + \mathcal{O}(t)$ at short times {and integrating on \(t\), the leading $1/z^2$} term governs the $|z| \to \infty$ asymptotic behavior:
\begin{equation}
\label{eq:asymptotic}
{\lim_{|z| \to \infty} z^2 \tilde{\sigma}_{xy}(z)} = \frac{i}{\hbar V} \langle [\hat{J}_x, \hat{J}_y] \rangle \equiv \mathcal{C}.
\end{equation}

Assuming causality, $z \tilde{\sigma}_{xy}(z)$ is analytic in the upper half-plane. Integrating over a contour enclosing this half-plane, Cauchy's theorem dictates that the real-axis integral is precisely canceled by that from the infinite semicircle, which equals $i \pi \mathcal{C}$:
\begin{equation}
\int_{-\infty}^{\infty} d\omega \, \omega \tilde{\sigma}_{xy}(\omega) =i \int_{-\infty}^{\infty} d\omega \, \omega \operatorname{Im}\tilde{\sigma}_{xy}(\omega)
= -i \pi \mathcal{C}.
\end{equation}
Here the first equality follows from the symmetry relation $\tilde{\sigma}_{xy}(-\omega)=\tilde{\sigma}_{xy}^*(\omega)$, which ensures that the integral of $\omega \operatorname{Re}\tilde{\sigma}_{xy}(\omega)$ vanishes.


Restricting the integral to positive frequencies and substituting $\mathcal{C}$, we arrive at the exact sum rule:
\begin{equation}
\int_0^\infty d\omega \, \omega \operatorname{Im}\tilde{\sigma}_{xy}(\omega) = -\frac{\pi}{2} \frac{i}{\hbar V} \langle [\hat{J}_x, \hat{J}_y] \rangle. \label{eq:sum_rule}
\end{equation}
Equation~(\ref{eq:sum_rule}) expresses the first frequency moment of antisymmetric optical Hall absorption in terms of an equal-time current commutator. Although closely related to earlier Hall-conductivity sum rules, this form is especially useful for optical probes because it directly constrains chirality-resolved absorption. Its value in the present work lies in the fact that the commutator can be evaluated exactly in the two topological settings considered below.

To elucidate this sum rule, consider an $N$-particle system governed by:
\begin{equation}
\hat{H} = \sum_{j=1}^{N} \left[ \frac{(\hat{\boldsymbol{p}}_j - q \boldsymbol{A}(\hat{\boldsymbol{r}}_j))^2}{2m} + V(\hat{\boldsymbol{r}}_j) + \hat{H}_{\text{SOC}}(\hat{\boldsymbol{r}}_j, \hat{\boldsymbol{p}}_j) \right],
\end{equation}
where $\boldsymbol{A}$ is the vector potential and $\hat{H}_{\text{SOC}}$ is the spin-orbit coupling. The total current operator is $\hat{\boldsymbol{J}} = \frac{q}{m} \sum_j (\hat{\boldsymbol{p}}_j - q\boldsymbol{A} + \dots)$. In the absence of an external magnetic field and spin-orbit coupling, the velocity components commute, and the right-hand side of Eq.~(\ref{eq:sum_rule}) vanishes. Conversely, in a uniform magnetic field $\boldsymbol{B} = \nabla \times \boldsymbol{A}$ with negligible spin-orbit coupling, one has $[\hat{v}_{j,x}, \hat{v}_{j,y}] = i\hbar q B_{z} / m^2$. The commutator evaluates to $[\hat{J}_x, \hat{J}_y] = i\hbar q^3 B_{z} N / m^2$, and the sum rule becomes 
\begin{equation}
\int_0^\infty d\omega \, \omega \operatorname{Im}\tilde{\sigma}_{xy}(\omega) = \frac{\pi n q^3 B_z}{2 m^2}. \label{eq:unifrom_magnetic_field}
\end{equation}
where $n = N/V$ is the carrier density. We now turn to the zero-field moir\'e continuum case, where the same formal relation instead forces the first moment to vanish exactly.

\textit{Zero magnetic field limit.}---We first investigate the continuum model for rhombohedral twisted bilayer $\mathrm{MoTe_2}$ ($\theta \approx 1.9^\circ$) using parameters fitted from density functional theory (DFT)~\cite{xuMultipleChernBands2024}. At low energies, the valley and spin degrees of freedom are locked, and time-reversal symmetry relates the $(\uparrow, K)$ and $(\downarrow, -K)$ sectors. Focusing on the $(\uparrow, K)$ valley, the Hamiltonian for the valence holes is defined as $H_{\text{hole}}(\boldsymbol{k}) = - H^*_{\text{elec}}(\boldsymbol{k})$, yielding: 
\begin{equation}
\hat{H}_{\text{hole}, (\uparrow, K)} = 
\begin{pmatrix}
\frac{(\hat{\boldsymbol{k}}-\boldsymbol{\kappa}_+)^2}{2 m^*} - \Delta_+(\boldsymbol{r}) & -\Delta_T^{\dagger}(\boldsymbol{r}) \\
-\Delta_T(\boldsymbol{r}) & \frac{(\hat{\boldsymbol{k}}-\boldsymbol{\kappa}_-)^2}{2 m^*} - \Delta_-(\boldsymbol{r})
\end{pmatrix}.
\end{equation}
The diagonal terms $\Delta_{\pm}(\boldsymbol{r})$ represent the intralayer moir\'e potential for the top and bottom layers, while $\Delta_T(\boldsymbol{r})$ describes the interlayer moir\'e tunneling. These are parameterized as:
\begin{align}
\Delta_{\pm}(\boldsymbol{r}) &= 2V_1 \sum_{j=1}^{3} \cos(\boldsymbol{g}^{(1)}_{j} \cdot \boldsymbol{r} \pm \phi_1) + 2V_2 \sum_{j=1}^{3} \cos(\boldsymbol{g}^{(2)}_{j} \cdot \boldsymbol{r}), \\
\Delta_T(\boldsymbol{r}) &= w_1 \sum_{j=1}^{3} e^{-i \boldsymbol{q}^{(1)}_{j} \cdot \boldsymbol{r}} + w_2 \sum_{j=1}^{3} e^{-i \boldsymbol{q}^{(2)}_{j} \cdot \boldsymbol{r}}.
\end{align}
The momentum offset between the parabolic band bottoms of the two layers is $|\boldsymbol{\kappa}_+ - \boldsymbol{\kappa}_-| = |\boldsymbol{K}|\theta$. The reciprocal lattice vectors $\{\boldsymbol{g}^{(n)}_{j}\}$ denote intra-layer momentum transfers for the nearest ($n=1$) and next-nearest ($n=2$) harmonics, while $\{\boldsymbol{q}^{(n)}_{j}\}$ denotes the interlayer momentum transfer vectors. 

We adopt standard phenomenological parameters to match DFT calculations~\cite{xuMultipleChernBands2024}: $V_1=2.4\,\mathrm{meV}$ and $V_2=1.0\,\mathrm{meV}$ control the amplitude of the spatial modulations (with phase $\phi_1=-90^\circ$). The interlayer terms $w_1=-5.8\,\mathrm{meV}$ and $w_2=2.8\,\mathrm{meV}$ parameterize the intra- and inter-sublattice tunneling strengths, respectively, with an effective mass of $m^*=0.62\,m_e$. The chemical potential is set to strictly occupy the lowest hole band, which carries a Chern number $C=-1$. Interestingly, in this model the top five valence bands near the Fermi level all carry Chern number $C=-1$.

Crucially, the continuum Hamiltonian contains a kinetic term that depends only on momentum $\hat{\boldsymbol{p}}$ and moir\'e potentials that depend only on position $\hat{\boldsymbol{r}}$. As a result, the velocity operator is strictly proportional to $\hat{\boldsymbol{p}}$, and the bare current components commute exactly ($\langle [\hat{J}_x, \hat{J}_y] \rangle = 0$). Equation~\eqref{eq:sum_rule} therefore predicts a rigorously vanishing first moment. Figure~\ref{fig:1}(b) displays the antisymmetric conductivity $\mathrm{Im} \tilde{\sigma}_{xy}(\omega)$, revealing a prominent positive peak (transitions from the occupied band to the lowest unoccupied band) perfectly compensated by negative peaks at higher frequencies.

\begin{figure}[htbp]
\includegraphics[width=1\linewidth]{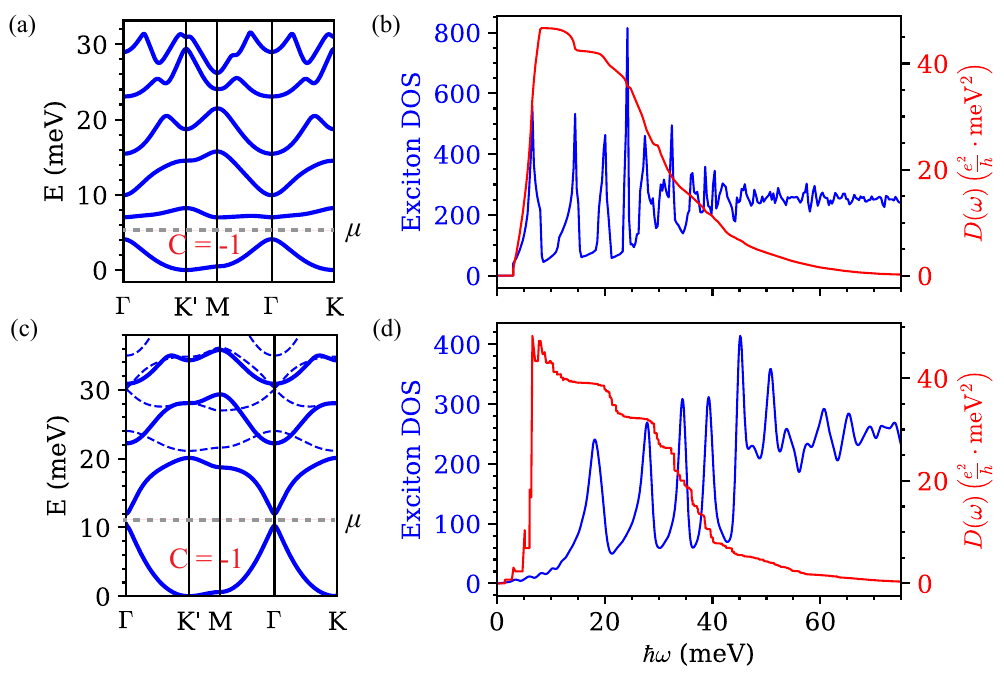}
\caption{\label{fig:2} (a) Non-interacting band structure of twisted MoTe$_2$ ($\theta=1.9^\circ$) in the hole basis. The gray dashed line indicates chemical potential $\mu$. (b) Cutoff-dependent first moment $D(\Omega)$ (red, right) and optical absorption spectrum (blue, left), normalized per eV per moir\'e unit cell. As $\Omega\to\infty$, $D(\Omega)\to 0$. (c) Self-consistent Hartree-Fock band structure. (d) Same as (b), computed from the interacting optical conductivity via the Bethe-Salpeter equation.}
\end{figure}

To verify the sum rule numerically, we compute the cutoff-dependent first moment:
\begin{equation}
D(\Omega) \equiv \int_{0}^{\Omega} d\omega \, \omega \, \mathrm{Im}\,\tilde{\sigma}_{xy}(\omega).
\end{equation}
As shown in Fig.~\ref{fig:2}(b), $D(\Omega)$ converges to zero as $\Omega \to \infty$, in exact agreement with the zero-field sum rule. However, $D(\Omega)$ remains sizable for cutoffs below $\sim 30\,\mathrm{meV}$---the energy window spanning the five consecutive $C=-1$ bands. 
If the optical probe is restricted to this range, the response mimics that of a system in a strong external magnetic field, where the Chern-band structure resembles a sequence of Landau levels. This apparent
low-energy ``violation" of the full sum rule reflects the noncommutative geometry of the projected low-energy subspace. The underlying zero-field character is recovered only when the integration cutoff extends beyond this topological low-energy manifold, allowing higher-band transitions to cancel the low-frequency Hall absorption exactly.

To ensure this result holds beyond the single-particle picture, we incorporate electron-electron interactions using the Hartree-Fock (HF) mean-field approximation. We assume a spontaneously valley-polarized $(\uparrow, K)$ ground state, consistent with experimental observations~\cite{kangEvidenceFractionalQuantum2024}. The interaction is modeled by a dual-gate screened Coulomb potential:
\begin{equation}
V(\boldsymbol{q}) = \frac{2\pi e^2}{\epsilon |\boldsymbol{q}|} \tanh(|\boldsymbol{q}|d),
\end{equation}
where $\epsilon=20$ is set to be the background dielectric constant and $d=20\,\mathrm{nm}$ is the distance to the metallic screening gates. To compute the interacting optical conductivity, we start from the HF band structure [Fig.~\ref{fig:2}(c)] and solve the Bethe-Salpeter equation (BSE) to include electron-hole attraction in the excitonic channels (see Supplemental Material (SM) for details~~\footnote{\label{fn}See Supplemental Material at http://link.aps.org/supplemental/xxx, for more details about optical response calculations in moir\'e continuum model and Hofstadter model, the correlation between optical weight and band mixing, and the connection to spontaneous Chern insulator, which includes Refs.~\cite{onishiFundamentalBoundTopological2023, qiuQuantumGeometryProbed2024, xuInterplayTopologyCorrelations2025, wuExcitonBandStructure2015, spantonObservationFractionalChern2018, wangChiralSuperconductivityFractional2025}.} )~\cite{wuExcitonBandStructure2015, qiuQuantumGeometryProbed2024}. The resulting interacting cutoff-dependent first moment [Fig.~\ref{fig:2}(d)] mirrors the non-interacting behavior: it initially rises to capture excitations across the lowest gap, and subsequently decreases to converge identically to zero at high frequencies. This explicitly confirms the robustness of the vanishing first-moment sum rule in the presence of interactions and excitonic effects.

\begin{figure}[htbp]
\includegraphics[width=1\linewidth]{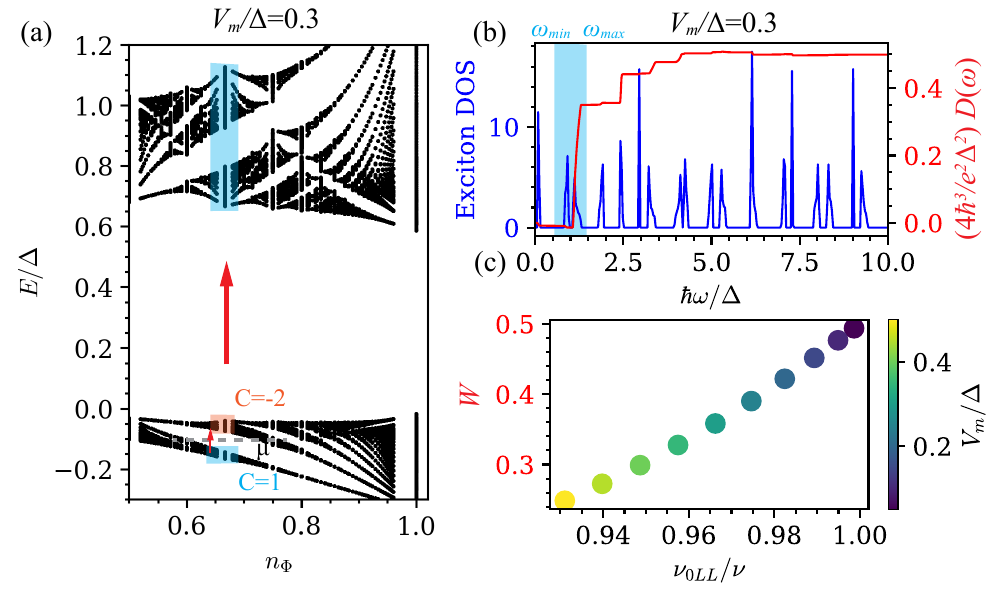}
\caption{\label{fig:4}(a) Hofstadter energy spectrum versus magnetic flux density $n_\Phi = \Phi/\Phi_0$ at $V_m/\Delta=0.3$.  At $n_\Phi=2/3$, the 0LL splits into two subbands (total $C=-1$).  The chemical potential $\mu$ (gray dashed) occupies only the lower subband of the split 0LL.$\Delta = \hbar\omega_c$ is the clean gap. (b) Exciton DOS (blue, left) and cumulative first moment (red, right) at $n_\Phi=2/3$, normalized to approach total filling $\nu$ as $\omega \to \infty$. The shaded window $[\omega_{\min}, \omega_{\max}]$ captures 0LL to 1LL transitions. (c) Partial sum rule contribution $W$ (interband weight in shaded window, see main text for detail) versus the 0LL ground-state fractional weight, $\nu_{0\mathrm{LL}}/\nu$, under varying $V_m/\Delta$ (color scale).}
\end{figure}

\textit{Uniform magnetic field limit.}---We next consider an electron system in a uniform magnetic field $\boldsymbol{B}=-B\hat{z}$ with negligible spin-orbit coupling. 
In this case the current commutator evaluates to $\langle [\hat{J}_x, \hat{J}_y] \rangle = i \hbar e^3 N B / m^2$ with electron charge to be \(-e\). We test this robust spectral weight constraint [Eq.~\eqref{eq:sum_rule}] using a periodic potential where the Hamiltonian assembles a Hofstadter model [Fig.~\ref{fig:1}(c)]~\cite{spantonObservationFractionalChern2018}:
\begin{equation}
\hat{H}_0 = \frac{(\hat{\boldsymbol{p}} +e\boldsymbol{A}(\hat{\boldsymbol{r}}))^2}{2m} + 2V_m \sum_{j=1,3,5} \cos(\boldsymbol{b}_j \cdot \hat{\boldsymbol{r}} + \phi),
\label{eq:H0_LL}
\end{equation}
where $V_m$ parameterizes the triangular-lattice potential ($\boldsymbol{b}_5 = -\boldsymbol{b}_1 - \boldsymbol{b}_3$, $\phi=0$). We assume a commensurate flux $\Phi$ per primitive cell ($A_{\mathrm{uc}} = \frac{\sqrt{3}}{2}a^2$):
\begin{equation}
\frac{\Phi}{\Phi_0} = \frac{B A_{\mathrm{uc}}}{\Phi_0} = \frac{p}{q}, \quad p,q \in \mathbb{Z}, \quad \gcd(p,q)=1,
\label{eq:flux_rational}
\end{equation}
for which the periodic potential splits the $n$-th Landau level (LL) into $p$ subbands [Fig.~\ref{fig:4}(a)]. We focus on $p/q = 2/3$ at filling $\nu=1/2$, such that only the lower subband ($C=1$) of the split 0LL is occupied.

In the clean limit ($V_m \to 0$), intra-0LL transitions are forbidden, and the response is a pure cyclotron resonance,
\begin{equation}
\operatorname{Im}\tilde{\sigma}_{xy}(\omega) = \frac{\pi n e^2}{2m} \left[ \delta(\omega - \omega_c) - \delta(\omega + \omega_c) \right],
\end{equation}
with cyclotron gap $\Delta = \hbar\omega_c$. Moreover, this cyclotron response remains unchanged even in the presence of electron-electron interactions, as guaranteed by Kohn's theorem~\cite{kohnCyclotronResonanceHaasvan1961}. 

For finite $V_m/\Delta$, however, exact translation symmetry is broken and higher-LL admixture is induced. As a result, $\operatorname{Im}\tilde{\sigma}_{xy}(\omega)$ becomes non-zero at low frequencies ($\hbar \omega \ll \Delta$) within the broadened 0LL manifold, while the primary cyclotron peak broadens and additional features of excitations appear near higher harmonics ($\omega \approx n \omega_c$), as shown in Fig.~\ref{fig:1}(d).

These different frequency regimes contribute to static topological and dynamic optical bounds in drastically different ways. Because the Kramers-Kronig relation weights the spectrum by $1/\omega$, both low- and high-frequency peaks contribute significantly to the dc Hall response,
\begin{equation}
\operatorname{Re}\tilde{\sigma}_{xy}(\omega=0) = \frac{2}{\pi} \int_0^\infty \frac{\operatorname{Im}\tilde{\sigma}_{xy}(\omega')}{\omega'} d\omega',
\label{eq:KK_sum_rule}
\end{equation}
yielding $C=1$ and $C=-2$ for the lower and upper subbands, respectively, for the potential considered here. In stark contrast, our derived first moment [Eq.~\eqref{eq:unifrom_magnetic_field}] is weighted by $\omega$ and is fixed entirely by the filling and magnetic field. Consequently, the low-frequency intra-LL peak contributes negligibly, while higher harmonics carry substantial weight [Fig.~\ref{fig:4}(b)]. 

To quantify this spectral redistribution, we define a partial spectral weight $W$ around the primary cyclotron transition:
\begin{equation}
W = \frac{4\hbar^3}{e^2 \Delta^2} \int_{\omega_{\min}}^{\omega_{\max}} d\omega \, \omega \operatorname{Im}\tilde{\sigma}_{xy}(\omega).
\end{equation}
In the clean limit ($V_m=0$), one has $W = \nu$. As $V_m$ increases, $W$ decreases, reflecting the transfer of spectral weight to higher harmonics. This reduction serves as an optical signature of LL mixing, which we quantify via the ground-state projection onto the ideal 0LL: 
\begin{equation}
    Z = \nu_{0\mathrm{LL}}/\nu = N_e^{-1} \sum_i \langle \Psi | \hat{P}_{0\mathrm{LL}}^{(i)} | \Psi \rangle. 
\end{equation}
Physically, $\nu_{0\mathrm{LL}} < \nu$ implies that higher LL components admix into the 0LL, thereby suppressing the primary cyclotron transition and enhancing higher-frequency excitations. Figure~\ref{fig:4}(c) demonstrates that for weak mixing, the reduction in $W$ scales approximately linearly with the loss of 0LL projection. Pictorially, the reduction arises mainly from weight transfers to higher-frequency peaks by band mixing; a detailed analysis is included in Sec. C of SM~\footnotemark[\value{footnote}]. Although the proportionality coefficient is nonuniversal (typically of order unity), $W$ provides a direct optical proxy for the extent of LL mixing. 

\textit{Discussion and Conclusion}.---We have established an exact optical sum rule relating the first frequency moment of $\operatorname{Im}\tilde{\sigma}_{xy}(\omega)$ to the ground-state current commutator $\langle [\hat{J}_x,\hat{J}_y] \rangle$. Since $\operatorname{Im}\tilde{\sigma}_{xy}(\omega)$ governs circular dichroism, this result imposes a model-independent constraint on the chiral optical response of interacting topological phases.

For continuum systems with parabolic dispersion and no extra momentum-dependent coupling such as SOC , the bare current operators commute exactly, $[\hat{J}_x,\hat{J}_y]=0$ \footnote{The same conclusion applies to lattice models whose kinetic terms preserve unit-cell translation symmetry and do not contain density-dependent correlated hopping.}, so the net first moment vanishes. As a result, in zero-field systems where time-reversal symmetry is broken internally, such as skyrmion topological Hall systems~\cite{wuTopologicalInsulatorsTwisted2019}, Kondo lattices~\cite{martinItinerantElectronDrivenChiral2008, rosalesFrustratedMagnetismSpontaneous2019}, and spontaneous Chern insulators~\cite{raghuTopologicalMottInsulators2008, renOrbitalChernInsulator2021}, any low-frequency Hall absorption must be compensated by higher-frequency spectral weight of the opposite sign. This implies a necessary sign reversal in the circular-dichroism spectrum and provides a criterion for distinguishing internally generated topology from that induced by orbital magnetic fields.

In uniform magnetic fields, by contrast, the same moment acquires a universal nonzero value fixed by the magnetic field and filling. Although periodic potentials redistribute spectral weight through Landau-level (LL) mixing and inter-subband transitions, the full first moment remains exactly constrained. In this setting, the deviation of the partial spectral weight $W$ from its clean-limit value provides a direct optical measure of LL mixing, relevant to topological phase transitions in two-dimensional electron gases and moir\'e or nanofabricated superlattices~\cite{wuHubbardModelPhysics2018, ciorciaroKineticMagnetismTriangular2023, kometterHofstadterStatesReentrant2023, singhaTwoDimensionalMottHubbardElectrons2011, levinParticleHoleSymmetryPfaffian2007, yuMoireFractionalChern2025}.

An important limitation arises in relativistic Dirac systems such as pristine graphene, where the velocity operators are proportional to Pauli matrices and the infinite Dirac sea renders the spectral integral divergent. Extending the exact sum rule to such gapless relativistic systems therefore requires a careful regularization of the infinite-band contribution, which remains an open question.

Overall, our results place optical Hall absorption on similar footing to conventional longitudinal sum rules while highlighting a qualitative distinction between zero-field topological response generated internally and that induced by orbital magnetic coupling. More broadly, they identify the Hall first moment as an experimentally relevant probe of spectral compensation, LL mixing, and the microscopic origin of topological optical response in interacting quantum materials.

\begin{acknowledgments}
We thank Hui Zeng, Nianlong Zou, Nikolai Peshcherenko and Joel Moore for their valuable discussions. We also thank Michael Zaletel for sharing the code for building the multicomponent quantum Hall systems Hamiltonian. This work is supported by the National Key R\&D Program of China (Grant No. 2021YFA1401600), the National Natural Science Foundation of China (Grant No. 12474056), and the 2022 basic discipline top-notch students training program 2.0 research project (Grant No. 20222005). Y.Z. acknowledges the support from the National Natural Science Foundation Young Student Basic Research Project (Grant No. 124B1030). The work was carried out at National Supercomputer Center in Tianjin, and the calculations were performed on Tianhe new generation supercomputer. The high-performance computing platform of Peking University supported the computational resources.
\end{acknowledgments}

\bibliography{apssamp}

@article{bacProbingBerryCurvature2025,
  title = {Probing {{Berry Curvature}} in {{Magnetic Topological Insulators}} through {{Resonant Infrared Magnetic Circular Dichroism}}},
  author = {Bac, Seul-Ki and Le Mardel{\'e}, Florian and Wang, Jiashu and Ozerov, Mykhaylo and Yoshimura, Kota and Mohelsk{\'y}, Ivan and Sun, Xingdan and Piot, Benjamin A. and Wimmer, Stefan and Ney, Andreas and Orlova, Tatyana and Zhukovskyi, Maksym and Bauer, G{\"u}nther and Springholz, Gunther and Liu, Xinyu and Orlita, Milan and Park, Kyungwha and Hsu, Yi-Ting and Assaf, Badih A.},
  year = 2025,
  month = jan,
  journal = {Physical Review Letters},
  volume = {134},
  number = {1},
  pages = {016601},
  publisher = {American Physical Society},
  doi = {10.1103/PhysRevLett.134.016601}
}

@article{caiSignaturesFractionalQuantum2023,
  title = {Signatures of Fractional Quantum Anomalous {{Hall}} States in Twisted {{MoTe2}}},
  author = {Cai, Jiaqi and Anderson, Eric and Wang, Chong and Zhang, Xiaowei and Liu, Xiaoyu and Holtzmann, William and Zhang, Yinong and Fan, Fengren and Taniguchi, Takashi and Watanabe, Kenji and Ran, Ying and Cao, Ting and Fu, Liang and Xiao, Di and Yao, Wang and Xu, Xiaodong},
  year = 2023,
  month = oct,
  journal = {Nature},
  volume = {622},
  number = {7981},
  pages = {63--68},
  publisher = {Nature Publishing Group},
  issn = {1476-4687},
  doi = {10.1038/s41586-023-06289-w},
  copyright = {2023 The Author(s), under exclusive licence to Springer Nature Limited},
  langid = {english}
}

@article{ciorciaroKineticMagnetismTriangular2023,
  title = {Kinetic Magnetism in Triangular Moir\'e Materials},
  author = {Ciorciaro, L. and Smole{\'n}ski, T. and Morera, I. and Kiper, N. and Hiestand, S. and Kroner, M. and Zhang, Y. and Watanabe, K. and Taniguchi, T. and Demler, E. and {\.I}mamo{\u g}lu, A.},
  year = 2023,
  month = nov,
  journal = {Nature},
  volume = {623},
  number = {7987},
  pages = {509--513},
  publisher = {Nature Publishing Group},
  issn = {1476-4687},
  doi = {10.1038/s41586-023-06633-0},
  copyright = {2023 The Author(s)},
  langid = {english}
}

@article{drewSumRuleOptical1997,
  title = {Sum {{Rule}} for the {{Optical Hall Angle}}},
  author = {Drew, H. D. and Coleman, P.},
  year = 1997,
  month = feb,
  journal = {Physical Review Letters},
  volume = {78},
  number = {8},
  pages = {1572--1575},
  publisher = {American Physical Society},
  doi = {10.1103/PhysRevLett.78.1572}
}

@article{dziomObservationUniversalMagnetoelectric2017,
  title = {Observation of the Universal Magnetoelectric Effect in a {{3D}} Topological Insulator},
  author = {Dziom, V. and Shuvaev, A. and Pimenov, A. and Astakhov, G. V. and Ames, C. and Bendias, K. and B{\"o}ttcher, J. and Tkachov, G. and Hankiewicz, E. M. and Br{\"u}ne, C. and Buhmann, H. and Molenkamp, L. W.},
  year = 2017,
  month = may,
  journal = {Nature Communications},
  volume = {8},
  number = {1},
  pages = {15197},
  publisher = {Nature Publishing Group},
  issn = {2041-1723},
  doi = {10.1038/ncomms15197},
  copyright = {2017 The Author(s)},
  langid = {english}
}

@misc{ghoshProbingQuantumGeometry2024,
  title = {Probing Quantum Geometry through Optical Conductivity and Magnetic Circular Dichroism},
  author = {Ghosh, Barun and Onishi, Yugo and Xu, Su-Yang and Lin, Hsin and Fu, Liang and Bansil, Arun},
  year = 2024,
  month = jan,
  number = {arXiv:2401.09689},
  eprint = {2401.09689},
  primaryclass = {cond-mat},
  publisher = {arXiv},
  doi = {10.48550/arXiv.2401.09689},
  archiveprefix = {arXiv}
}

@book{giulianiQuantumTheoryElectron2005,
  title = {Quantum {{Theory}} of the {{Electron Liquid}}},
  author = {Giuliani, Gabriele and Vignale, Giovanni},
  year = 2005,
  publisher = {Cambridge University Press},
  address = {Cambridge},
  doi = {10.1017/CBO9780511619915},
  isbn = {978-0-521-52796-5}
}

@article{haldaneBerryCurvatureFermi2004,
  title = {Berry {{Curvature}} on the {{Fermi Surface}}: {{Anomalous Hall Effect}} as a {{Topological Fermi-Liquid Property}}},
  shorttitle = {Berry {{Curvature}} on the {{Fermi Surface}}},
  author = {Haldane, F. D. M.},
  year = 2004,
  month = nov,
  journal = {Physical Review Letters},
  volume = {93},
  number = {20},
  pages = {206602},
  publisher = {American Physical Society},
  doi = {10.1103/PhysRevLett.93.206602}
}

@article{jungwirthAnomalousHallEffect2002,
  title = {Anomalous {{Hall Effect}} in {{Ferromagnetic Semiconductors}}},
  author = {Jungwirth, T. and Niu, Qian and MacDonald, A. H.},
  year = 2002,
  month = may,
  journal = {Physical Review Letters},
  volume = {88},
  number = {20},
  pages = {207208},
  publisher = {American Physical Society},
  doi = {10.1103/PhysRevLett.88.207208}
}

@article{kangEvidenceFractionalQuantum2024,
  title = {Evidence of the Fractional Quantum Spin {{Hall}} Effect in Moir\'e {{MoTe2}}},
  author = {Kang, Kaifei and Shen, Bowen and Qiu, Yichen and Zeng, Yihang and Xia, Zhengchao and Watanabe, Kenji and Taniguchi, Takashi and Shan, Jie and Mak, Kin Fai},
  year = 2024,
  month = mar,
  journal = {Nature},
  pages = {1--5},
  publisher = {Nature Publishing Group},
  issn = {1476-4687},
  doi = {10.1038/s41586-024-07214-5},
  copyright = {2024 The Author(s), under exclusive licence to Springer Nature Limited},
  langid = {english}
}

@article{katsufujiSpectralWeightTransfer1995,
  title = {Spectral {{Weight Transfer}} of the {{Optical Conductivity}} in {{Doped Mott Insulators}}},
  author = {Katsufuji, T. and Okimoto, Y. and Tokura, Y.},
  year = 1995,
  month = nov,
  journal = {Physical Review Letters},
  volume = {75},
  number = {19},
  pages = {3497--3500},
  publisher = {American Physical Society},
  doi = {10.1103/PhysRevLett.75.3497}
}

@article{klitzingNewMethodHighAccuracy1980,
  title = {New {{Method}} for {{High-Accuracy Determination}} of the {{Fine-Structure Constant Based}} on {{Quantized Hall Resistance}}},
  author = {v. Klitzing, K. and Dorda, G. and Pepper, M.},
  year = 1980,
  month = aug,
  journal = {Physical Review Letters},
  volume = {45},
  number = {6},
  pages = {494--497},
  publisher = {American Physical Society},
  doi = {10.1103/PhysRevLett.45.494}
}

@article{kohnCyclotronResonanceHaasvan1961,
  title = {Cyclotron {{Resonance}} and de {{Haas-van Alphen Oscillations}} of an {{Interacting Electron Gas}}},
  author = {Kohn, Walter},
  year = 1961,
  month = aug,
  journal = {Physical Review},
  volume = {123},
  number = {4},
  pages = {1242--1244},
  publisher = {American Physical Society},
  doi = {10.1103/PhysRev.123.1242}
}

@article{kometterHofstadterStatesReentrant2023,
  title = {Hofstadter States and Re-Entrant Charge Order in a Semiconductor Moir\'e Lattice},
  author = {Kometter, Carlos R. and Yu, Jiachen and Devakul, Trithep and Reddy, Aidan P. and Zhang, Yang and Foutty, Benjamin A. and Watanabe, Kenji and Taniguchi, Takashi and Fu, Liang and Feldman, Benjamin E.},
  year = 2023,
  month = dec,
  journal = {Nature Physics},
  volume = {19},
  number = {12},
  pages = {1861--1867},
  publisher = {Nature Publishing Group},
  issn = {1745-2481},
  doi = {10.1038/s41567-023-02195-0},
  copyright = {2023 The Author(s), under exclusive licence to Springer Nature Limited},
  langid = {english}
}

@article{langeMagnetoopticalSumRules1999,
  title = {Magneto-Optical {{Sum Rules Close}} to the {{Mott Transition}}},
  author = {Lange, Ekkehard and Kotliar, Gabriel},
  year = 1999,
  month = feb,
  journal = {Physical Review Letters},
  volume = {82},
  number = {6},
  pages = {1317--1320},
  publisher = {American Physical Society},
  doi = {10.1103/PhysRevLett.82.1317}
}

@article{langeMemoryfunctionApproachHall1997,
  title = {Memory-Function Approach to the {{Hall}} Constant in Strongly Correlated Electron Systems},
  author = {Lange, Ekkehard},
  year = 1997,
  month = feb,
  journal = {Physical Review B},
  volume = {55},
  number = {6},
  pages = {3907--3928},
  publisher = {American Physical Society},
  doi = {10.1103/PhysRevB.55.3907}
}

@article{levinParticleHoleSymmetryPfaffian2007,
  title = {Particle-{{Hole Symmetry}} and the {{Pfaffian State}}},
  author = {Levin, Michael and Halperin, Bertrand I. and Rosenow, Bernd},
  year = 2007,
  month = dec,
  journal = {Physical Review Letters},
  volume = {99},
  number = {23},
  pages = {236806},
  publisher = {American Physical Society},
  doi = {10.1103/PhysRevLett.99.236806}
}

@article{martinItinerantElectronDrivenChiral2008,
  title = {Itinerant {{Electron-Driven Chiral Magnetic Ordering}} and {{Spontaneous Quantum Hall Effect}} in {{Triangular Lattice Models}}},
  author = {Martin, Ivar and Batista, C. D.},
  year = 2008,
  month = oct,
  journal = {Physical Review Letters},
  volume = {101},
  number = {15},
  pages = {156402},
  publisher = {American Physical Society},
  doi = {10.1103/PhysRevLett.101.156402}
}

@article{moligniniProbingChernNumber2023,
  title = {Probing {{Chern}} Number by Opacity and Topological Phase Transition by a Nonlocal {{Chern}} Marker},
  author = {Molignini, Paolo and Lapierre, Bastien and Chitra, Ramasubramanian and Chen, Wei},
  year = 2023,
  month = aug,
  journal = {SciPost Physics Core},
  volume = {6},
  number = {3},
  pages = {059},
  issn = {2666-9366},
  doi = {10.21468/SciPostPhysCore.6.3.059},
  langid = {english}
}

@article{okadaTerahertzSpectroscopyFaraday2016,
  title = {Terahertz Spectroscopy on {{Faraday}} and {{Kerr}} Rotations in a Quantum Anomalous {{Hall}} State},
  author = {Okada, Ken N. and Takahashi, Youtarou and Mogi, Masataka and Yoshimi, Ryutaro and Tsukazaki, Atsushi and Takahashi, Kei S. and Ogawa, Naoki and Kawasaki, Masashi and Tokura, Yoshinori},
  year = 2016,
  month = jul,
  journal = {Nature Communications},
  volume = {7},
  number = {1},
  pages = {12245},
  publisher = {Nature Publishing Group},
  issn = {2041-1723},
  doi = {10.1038/ncomms12245},
  copyright = {2016 The Author(s)},
  langid = {english}
}

@misc{onishiFundamentalBoundTopological2023,
  title = {Fundamental Bound on Topological Gap},
  author = {Onishi, Yugo and Fu, Liang},
  year = 2023,
  month = oct,
  number = {arXiv:2306.00078},
  eprint = {2306.00078},
  primaryclass = {cond-mat},
  publisher = {arXiv},
  doi = {10.48550/arXiv.2306.00078},
  archiveprefix = {arXiv}
}

@misc{onishiQuantumWeight2024,
  title = {Quantum Weight},
  author = {Onishi, Yugo and Fu, Liang},
  year = 2024,
  month = jan,
  number = {arXiv:2401.13847},
  eprint = {2401.13847},
  primaryclass = {cond-mat},
  publisher = {arXiv},
  doi = {10.48550/arXiv.2401.13847},
  archiveprefix = {arXiv}
}

@misc{onishiTopologicalBoundStructure2024,
  title = {Topological Bound on Structure Factor},
  author = {Onishi, Yugo and Fu, Liang},
  year = 2024,
  month = oct,
  number = {arXiv:2406.18654},
  eprint = {2406.18654},
  publisher = {arXiv},
  doi = {10.48550/arXiv.2406.18654},
  archiveprefix = {arXiv}
}

@misc{onishiUniversalRelationEnergy2024,
  title = {Universal Relation between Energy Gap and Dielectric Constant},
  author = {Onishi, Yugo and Fu, Liang},
  year = 2024,
  month = feb,
  number = {arXiv:2401.04180},
  eprint = {2401.04180},
  primaryclass = {cond-mat},
  publisher = {arXiv},
  doi = {10.48550/arXiv.2401.04180},
  archiveprefix = {arXiv}
}

@article{onodaTopologicalNatureAnomalous2002,
  title = {Topological {{Nature}} of {{Anomalous Hall Effect}} in {{Ferromagnets}}},
  author = {Onoda, Masaru and Nagaosa, Naoto},
  year = 2002,
  month = jan,
  journal = {Journal of the Physical Society of Japan},
  volume = {71},
  number = {1},
  pages = {19--22},
  issn = {0031-9015, 1347-4073},
  doi = {10.1143/JPSJ.71.19},
  langid = {english}
}

@misc{qiuQuantumGeometryProbed2024,
  title = {Quantum {{Geometry Probed}} by {{Chiral Excitonic Optical Response}} of {{Chern Insulators}}},
  author = {Qiu, Wen-Xuan and Wu, Fengcheng},
  year = 2024,
  month = jul,
  number = {arXiv:2407.03317},
  eprint = {2407.03317},
  publisher = {arXiv},
  doi = {10.48550/arXiv.2407.03317},
  archiveprefix = {arXiv}
}

@article{raghuTopologicalMottInsulators2008,
  title = {Topological {{Mott Insulators}}},
  author = {Raghu, S. and Qi, Xiao-Liang and Honerkamp, C. and Zhang, Shou-Cheng},
  year = 2008,
  month = apr,
  journal = {Physical Review Letters},
  volume = {100},
  number = {15},
  pages = {156401},
  publisher = {American Physical Society},
  doi = {10.1103/PhysRevLett.100.156401}
}

@article{reicheUeberZahlDispersionselektronen1925,
  title = {{\"Uber die Zahl der Dispersionselektronen, die einem station\"aren Zustand zugeordnet sind}},
  author = {Reiche, F. and Thomas, W.},
  year = 1925,
  month = dec,
  journal = {Zeitschrift f\"ur Physik},
  volume = {34},
  number = {1},
  pages = {510--525},
  issn = {0044-3328},
  doi = {10.1007/BF01328494},
  langid = {ngerman}
}

@article{renOrbitalChernInsulator2021,
  title = {Orbital {{Chern Insulator}} and {{Quantum Phase Diagram}} of a {{Kagome Electron System}} with {{Half-Filled Flat Bands}}},
  author = {Ren, Yafei and Jiang, Hong-Chen and Qiao, Zhenhua and Sheng, D. N.},
  year = 2021,
  month = mar,
  journal = {Physical Review Letters},
  volume = {126},
  number = {11},
  pages = {117602},
  publisher = {American Physical Society},
  doi = {10.1103/PhysRevLett.126.117602}
}

@article{rosalesFrustratedMagnetismSpontaneous2019,
  title = {From Frustrated Magnetism to Spontaneous {{Chern}} Insulators},
  author = {Rosales, H. D. and Albarrac{\'i}n, F. A. G{\'o}mez and Pujol, P.},
  year = 2019,
  month = jan,
  journal = {Physical Review B},
  volume = {99},
  number = {3},
  pages = {035163},
  publisher = {American Physical Society},
  doi = {10.1103/PhysRevB.99.035163}
}

@article{rozenbergTransferSpectralWeight1996,
  title = {Transfer of Spectral Weight in Spectroscopies of Correlated Electron Systems},
  author = {Rozenberg, M. J. and Kotliar, G. and Kajueter, H.},
  year = 1996,
  month = sep,
  journal = {Physical Review B},
  volume = {54},
  number = {12},
  pages = {8452--8468},
  publisher = {American Physical Society},
  doi = {10.1103/PhysRevB.54.8452}
}

@article{singhaTwoDimensionalMottHubbardElectrons2011,
  title = {Two-{{Dimensional Mott-Hubbard Electrons}} in an {{Artificial Honeycomb Lattice}}},
  author = {Singha, A. and Gibertini, M. and Karmakar, B. and Yuan, S. and Polini, M. and Vignale, G. and Katsnelson, M. I. and Pinczuk, A. and Pfeiffer, L. N. and West, K. W. and Pellegrini, V.},
  year = 2011,
  month = jun,
  journal = {Science},
  volume = {332},
  number = {6034},
  pages = {1176--1179},
  publisher = {American Association for the Advancement of Science},
  doi = {10.1126/science.1204333}
}

@article{souzaDichroic$f$sumRule2008,
  title = {Dichroic \$f\$-Sum Rule and the Orbital Magnetization of Crystals},
  author = {Souza, Ivo and Vanderbilt, David},
  year = 2008,
  month = feb,
  journal = {Physical Review B},
  volume = {77},
  number = {5},
  pages = {054438},
  publisher = {American Physical Society},
  doi = {10.1103/PhysRevB.77.054438}
}

@article{spantonObservationFractionalChern2018,
  title = {Observation of Fractional {{Chern}} Insulators in a van Der {{Waals}} Heterostructure},
  author = {Spanton, Eric M. and Zibrov, Alexander A. and Zhou, Haoxin and Taniguchi, Takashi and Watanabe, Kenji and Zaletel, Michael P. and Young, Andrea F.},
  year = 2018,
  month = apr,
  journal = {Science},
  volume = {360},
  number = {6384},
  pages = {62--66},
  publisher = {American Association for the Advancement of Science},
  doi = {10.1126/science.aan8458}
}

@article{taoValleyCoherentQuantumAnomalous2024,
 title = {Valley-Coherent Quantum Anomalous Hall State in AB-Stacked ${\mathrm{MoTe}}_{2}/{\mathrm{W}\mathrm{S}\mathrm{e}}_{2}$ Bilayers},
  author = {Tao, Zui and Shen, Bowen and Jiang, Shengwei and Li, Tingxin and Li, Lizhong and Ma, Liguo and Zhao, Wenjin and Hu, Jenny and Pistunova, Kateryna and Watanabe, Kenji and Taniguchi, Takashi and Heinz, Tony F. and Mak, Kin Fai and Shan, Jie},
  journal = {Phys. Rev. X},
  volume = {14},
  issue = {1},
  pages = {011004},
  numpages = {8},
  year = {2024},
  month = {Jan},
  publisher = {American Physical Society},
  doi = {10.1103/PhysRevX.14.011004},
  url = {https://link.aps.org/doi/10.1103/PhysRevX.14.011004}
}

@article{tholeXrayCircularDichroism1992,
  title = {X-Ray Circular Dichroism as a Probe of Orbital Magnetization},
  author = {Thole, B. T. and Carra, P. and Sette, F. and {van der Laan}, G.},
  year = 1992,
  month = mar,
  journal = {Physical Review Letters},
  volume = {68},
  number = {12},
  pages = {1943--1946},
  publisher = {American Physical Society},
  doi = {10.1103/PhysRevLett.68.1943}
}

@article{uchidaOpticalSpectra$mathrmLa_2mathrmensuremathmathitx$$mathrmSr_mathitx$$mathrmCuO_4$1991,
 title = {Optical spectra of ${\mathrm{La}}_{2\mathrm{\ensuremath{-}}\mathit{x}}$${\mathrm{Sr}}_{\mathit{x}}$${\mathrm{CuO}}_{4}$: Effect of carrier doping on the electronic structure of the ${\mathrm{CuO}}_{2}$ plane},
  author = {Uchida, S. and Ido, T. and Takagi, H. and Arima, T. and Tokura, Y. and Tajima, S.},
  year = 1991,
  month = apr,
  journal = {Physical Review B},
  volume = {43},
  number = {10},
  pages = {7942--7954},
  publisher = {American Physical Society},
  doi = {10.1103/PhysRevB.43.7942}
}

@misc{vermaInstantaneousResponseQuantum2024,
  title = {Instantaneous {{Response}} and {{Quantum Geometry}} of {{Insulators}}},
  author = {Verma, Nishchhal and Queiroz, Raquel},
  year = 2024,
  month = mar,
  number = {arXiv:2403.07052},
  eprint = {2403.07052},
  publisher = {arXiv},
  doi = {10.48550/arXiv.2403.07052},
  archiveprefix = {arXiv}
}

@misc{wangChiralSuperconductivityFractional2025,
  title = {Chiral Superconductivity near a Fractional {{Chern}} Insulator},
  author = {Wang, Taige and Zaletel, Michael P.},
  year = 2025,
  month = jul,
  number = {arXiv:2507.07921},
  eprint = {2507.07921},
  primaryclass = {cond-mat},
  publisher = {arXiv},
  doi = {10.48550/arXiv.2507.07921},
  archiveprefix = {arXiv}
}

@article{wuExcitonBandStructure2015,
  title = {Exciton Band Structure of Monolayer \$\textbraceleft\textbackslash mathrm\textbraceleft{{MoS}}\textbraceright\textbraceright\_\textbraceleft 2\textbraceright\$},
  author = {Wu, Fengcheng and Qu, Fanyao and MacDonald, A. H.},
  year = 2015,
  month = feb,
  journal = {Physical Review B},
  volume = {91},
  number = {7},
  pages = {075310},
  publisher = {American Physical Society},
  doi = {10.1103/PhysRevB.91.075310}
}

@article{wuHubbardModelPhysics2018,
  title = {Hubbard {{Model Physics}} in {{Transition Metal Dichalcogenide Moir}}\textbackslash 'e {{Bands}}},
  author = {Wu, Fengcheng and Lovorn, Timothy and Tutuc, Emanuel and MacDonald, A. H.},
  year = 2018,
  month = jul,
  journal = {Physical Review Letters},
  volume = {121},
  number = {2},
  pages = {026402},
  publisher = {American Physical Society},
  doi = {10.1103/PhysRevLett.121.026402}
}

@article{wuQuantizedFaradayKerr2016,
  title = {Quantized {{Faraday}} and {{Kerr}} Rotation and Axion Electrodynamics of a {{3D}} Topological Insulator},
  author = {Wu, Liang and Salehi, M. and Koirala, N. and Moon, J. and Oh, S. and Armitage, N. P.},
  year = 2016,
  month = dec,
  journal = {Science},
  volume = {354},
  number = {6316},
  pages = {1124--1127},
  publisher = {American Association for the Advancement of Science},
  doi = {10.1126/science.aaf5541}
}

@article{wuTopologicalInsulatorsTwisted2019,
  title = {Topological {{Insulators}} in {{Twisted Transition Metal Dichalcogenide Homobilayers}}},
  author = {Wu, Fengcheng and Lovorn, Timothy and Tutuc, Emanuel and Martin, Ivar and MacDonald, A. H.},
  year = 2019,
  month = feb,
  journal = {Physical Review Letters},
  volume = {122},
  number = {8},
  pages = {086402},
  publisher = {American Physical Society},
  doi = {10.1103/PhysRevLett.122.086402}
}

@misc{xuMultipleChernBands2024,
  title = {Multiple {{Chern}} Bands in Twisted {{MoTe}}\$\_2\$ and Possible Non-{{Abelian}} States},
  author = {Xu, Cheng and Mao, Ning and Zeng, Tiansheng and Zhang, Yang},
  year = 2024,
  month = mar,
  number = {arXiv:2403.17003},
  eprint = {2403.17003},
  primaryclass = {cond-mat},
  publisher = {arXiv},
  doi = {10.48550/arXiv.2403.17003},
  archiveprefix = {arXiv}
}

@article{yuMoireFractionalChern2025,
  title = {Moir\textbackslash 'e Fractional {{Chern}} Insulators. {{IV}}. {{Fluctuation-driven}} Collapse in Multiband Exact Diagonalization Calculations on Rhombohedral Graphene},
  author = {Yu, Jiabin and {Herzog-Arbeitman}, Jonah and Kwan, Yves H. and Regnault, Nicolas and Bernevig, B. Andrei},
  year = 2025,
  month = aug,
  journal = {Physical Review B},
  volume = {112},
  number = {7},
  pages = {075110},
  publisher = {American Physical Society},
  doi = {10.1103/PhysRevB.112.075110}
}

@article{yuUniversalWilsonLoop2025,
  title = {Universal {{Wilson Loop Bound}} of {{Quantum Geometry}}},
  author = {Yu, Jiabin and {Herzog-Arbeitman}, Jonah and Bernevig, B. Andrei},
  year = 2025,
  month = aug,
  journal = {Physical Review Letters},
  volume = {135},
  number = {8},
  eprint = {2501.00100},
  primaryclass = {cond-mat},
  pages = {086401},
  issn = {0031-9007, 1079-7114},
  doi = {10.1103/mp2c-zzkt},
  archiveprefix = {arXiv}
}

@article{xuInterplayTopologyCorrelations2025,
  title = {Interplay between Topology and Correlations in the Second Moir\'e Band of Twisted Bilayer {{MoTe2}}},
  author = {Xu, Fan and Chang, Xumin and Xiao, Jiayong and Zhang, Yixin and Liu, Feng and Sun, Zheng and Mao, Ning and Peshcherenko, Nikolai and Li, Jiayi and Watanabe, Kenji and Taniguchi, Takashi and Tong, Bingbing and Lu, Li and Jia, Jinfeng and Qian, Dong and Shi, Zhiwen and Zhang, Yang and Liu, Xiaoxue and Jiang, Shengwei and Li, Tingxin},
  year = 2025,
  month = apr,
  journal = {Nature Physics},
  volume = {21},
  number = {4},
  pages = {542--548},
  publisher = {Nature Publishing Group},
  issn = {1745-2481},
  doi = {10.1038/s41567-025-02803-1},
  copyright = {2025 The Author(s), under exclusive licence to Springer Nature Limited},
  langid = {english}
}

@misc{passosElectronicBoundsMagnetic2025,
  title = {Electronic Bounds in Magnetic Crystals},
  author = {Passos, Daniel and Souza, Ivo},
  year = 2025,
  month = sep,
  number = {arXiv:2509.16121},
  eprint = {2509.16121},
  primaryclass = {cond-mat},
  publisher = {arXiv},
  doi = {10.48550/arXiv.2509.16121},
  archiveprefix = {arXiv}
}

@article{souzaOpticalBoundsManyelectron2025,
  title = {Optical Bounds on Many-Electron Localization},
  author = {Souza, Ivo and Martin, Richard M. and Stengel, Massimiliano},
  year = 2025,
  month = apr,
  journal = {SciPost Physics},
  volume = {18},
  number = {4},
  eprint = {2407.17908},
  primaryclass = {cond-mat},
  pages = {127},
  issn = {2542-4653},
  doi = {10.21468/SciPostPhys.18.4.127},
  archiveprefix = {arXiv}
}

\clearpage 
\appendix
\onecolumngrid
\newcommand{\beginsupplement}{%
        \setcounter{table}{0}
        \renewcommand{\thetable}{S\arabic{table}}%
        \setcounter{figure}{0}
        \renewcommand{\thefigure}{S\arabic{figure}}%
     }
\beginsupplement
\renewcommand{\theequation}{S\arabic{equation}}

\section*{
Supplementary Material for ``
Optical Hall absorption sum rule and spectral compensation in time-reversal-breaking moir\'e and Hofstadter systems''}
\twocolumngrid

\section{Optical response calculation in continuum models}

In the non-interacting limit, the optical conductivity of the insulating state is computed using the Kubo-Greenwood formula formulated within band theory~\cite{onishiFundamentalBoundTopological2023}:
\begin{equation}
\sigma_{\mu\nu}(\omega) = \frac{q^2}{\hbar} \int [d\boldsymbol{k}] \sum_{a \neq b} \frac{-i \varepsilon_{ba}(\boldsymbol{k}) \mathcal{A}^\mu_{ab}(\boldsymbol{k}) \mathcal{A}^\nu_{ba}(\boldsymbol{k})}{\hbar\omega - \varepsilon_{ba}(\boldsymbol{k}) + i\eta} f_{ab}(\boldsymbol{k}).
\label{eq:band-sigma}
\end{equation}
Here, $\varepsilon_{ba}(\boldsymbol{k}) = \varepsilon_b(\boldsymbol{k}) - \varepsilon_a(\boldsymbol{k})$ denotes the interband energy difference, $\mathcal{A}^\mu_{ab}(\boldsymbol{k}) = \langle u_a(\boldsymbol{k}) | i\partial_{k_\mu} | u_b(\boldsymbol{k}) \rangle$ is the Berry connection matrix element, $f_{ab}(\boldsymbol{k}) = f_a(\boldsymbol{k}) - f_b(\boldsymbol{k})$ is the difference in Fermi-Dirac occupation factors, and $\eta \to 0^+$ is a phenomenological broadening parameter.

To incorporate many-body interaction effects, we first determine the ground state $|G\rangle$ through a self-consistent Hartree-Fock (HF) analysis~\cite{qiuQuantumGeometryProbed2024, xuInterplayTopologyCorrelations2025}. The excited states and their corresponding energies, which capture the excitonic self-energy, are subsequently obtained by solving the lowest-order Bethe-Salpeter (BS) equation \cite{wuExcitonBandStructure2015}. The excitation wavefunction $|\chi\rangle$ is constructed as a linear combination of electron-hole pairs created across the HF gap:
\begin{equation}
|\chi\rangle = \sum_{\boldsymbol k,\lambda\ge2} z_{\boldsymbol k,\lambda}(\chi)\, f^{\dagger}_{\boldsymbol k\lambda} f_{\boldsymbol k 1}\,|G\rangle,
\end{equation}
where $\lambda=1$ explicitly labels the lowest, fully occupied hole band. The expansion coefficients $z_{\boldsymbol k,\lambda}(\chi)$ and the excitation energy $\varepsilon_\chi$ are determined by the eigenvalue problem:
\begin{equation}
\varepsilon_\chi\, z_{\boldsymbol k,\lambda}(\chi) = \sum_{\boldsymbol p,\lambda'} \mathcal H^{\lambda\lambda'}_{\boldsymbol k\boldsymbol p}\,z_{\boldsymbol p,\lambda'}(\chi),
\end{equation}
with the effective Hamiltonian matrix defined as:
\begin{equation}
\mathcal{H}_{\boldsymbol{k} \boldsymbol{p}}^{\lambda \lambda^{\prime}} = \left(E_{\boldsymbol{k}}^{\lambda}-E_{\boldsymbol{k}}^{1}\right) \delta_{\boldsymbol{k} \boldsymbol{p}} \delta_{\lambda \lambda^{\prime}} - \left(\tilde{V}_{\boldsymbol{p} \boldsymbol{k} \boldsymbol{p} \boldsymbol{k}}^{1 \lambda \lambda^{\prime} 1} - \tilde{V}_{\boldsymbol{k} \boldsymbol{p} \boldsymbol{p} \boldsymbol{k}}^{\lambda 1 \lambda^{\prime} 1}\right).
\end{equation}
In this expression, $E_{\boldsymbol{k}}^{\lambda}$ represents the HF band energy for band $\lambda$. The terms $\tilde{V}$ represent the screened Coulomb interaction $V(\boldsymbol{q})$ transformed from the plane-wave basis into the basis of HF band eigenstates, accounting for the relevant scattering processes between the excited electron and the hole.

The interacting optical response is then calculated by generalizing Eq.~\eqref{eq:band-sigma} to the many-body basis. Specifically, the single-particle Berry connection is replaced by the optical transition dipole element between the ground state and the exciton state $|\chi\rangle$, defined as $A_{\chi G}^{\mu} = \langle\chi|\hat{r}^\mu|G\rangle = \sum_{\boldsymbol k} \mathcal{A}^\mu_{\lambda 1}(\boldsymbol k)\,z^{\ast}_{\boldsymbol k,\lambda}(\chi)$. 

Finally, to ensure numerical convergence, our calculations utilize a $24\times24$ $\Gamma$-centered $\boldsymbol k$-mesh for both the HF self-consistency and the BS equations. We truncate the unoccupied bands at a maximum cutoff of $\lambda = 25$, beyond which the contribution to the spectral weight $\omega \, \mathrm{Im}\,\tilde{\sigma}_{xy}(\omega)$ becomes negligible.

\section{Optical response calculation in the Hofstadter model}

We consider the Hamiltonian for an electron in a parabolic band subjected to both a uniform magnetic field and a periodic triangular lattice potential \cite{spantonObservationFractionalChern2018, wangChiralSuperconductivityFractional2025}:
\begin{equation}
\hat{H}_0 = \frac{(\hat{\boldsymbol{p}} + e \boldsymbol{A}(\hat{\boldsymbol{r}}))^2}{2m} + 2V_m \sum_{j=1,3,5} \cos(\boldsymbol{b}_j \cdot \hat{\boldsymbol{r}} + \phi).
\end{equation}
To define a triangular lattice with lattice constant $a$, we choose the reciprocal lattice vectors $\boldsymbol{b}_1$, $\boldsymbol{b}_3$, and $\boldsymbol{b}_5$ to have equal magnitude $b = 4\pi/(\sqrt{3}a)$ and to be separated by angles of $120^\circ$. We align the coordinate system such that the $x$-axis is parallel to $\boldsymbol{b}_1$, while the $y$-axis is perpendicular to it. The reciprocal lattice vectors are therefore given by $\boldsymbol{b}_1 = (b, 0)$, $\boldsymbol{b}_3 = (-b/2, \sqrt{3}b/2)$, and $\boldsymbol{b}_5 = (-b/2, -\sqrt{3}b/2)$. In real space, the potential is periodic under translations by $\sqrt{3}a$ along the $x$ direction and by $a$ along the $y$ direction, naturally defining a rectangular extended unit cell with area $A_{\mathrm{rect}} = \sqrt{3}a^2$.

We assume that the charge carriers are electrons ($e>0$) and that the magnetic field points along the negative $\hat{z}$ direction, $\boldsymbol{B} = -B \hat{z}$ with $B>0$. Working in the Landau gauge $\boldsymbol{A} = (0, -Bx, 0)$, we span the unperturbed Hilbert space using the Landau-level (LL) basis $|n_{\text{LL}}, k_y\rangle$, where $n_{\text{LL}}$ denotes the LL index. To discretize the Hilbert space, we wrap the $y$ direction into a cylinder of circumference $L_y = N_y a$, where $N_y$ is a large integer. This imposes the momentum quantization condition $k_y = 2\pi j /(N_y a)$ for integer $j$.

Because the primary triangular unit cell, with area $A_{\mathrm{uc}} = \frac{\sqrt{3}}{2}a^2$, is pierced by a rational flux $\Phi/\Phi_0 = p/q$, the magnetic translation symmetry is reduced relative to that of the periodic triangular lattice potential. Specifically, our rectangular extended unit cell, with area $A_{\mathrm{rect}} = 2A_{\mathrm{uc}}$, contains $2p/q$ flux quanta, whereas magnetic Bloch theory applies only when the flux per magnetic unit cell is an integer. To satisfy this condition, we define a magnetic supercell spanning $q$ extended units along $x$ and one extended unit along $y$, so that it encloses exactly $2p$ flux quanta.

To diagonalize the Hamiltonian and obtain the folded Hofstadter subbands, we compute the matrix elements of the periodic potential in the unperturbed LL basis. For a single plane-wave component $e^{i\boldsymbol{q}\cdot\hat{\boldsymbol{r}}}$, where $\boldsymbol{q} = \pm \boldsymbol{b}_j$, the matrix element is
\begin{equation}
\langle n_{\text{LL}}', k_y' | e^{i\boldsymbol{q}\cdot\hat{\boldsymbol{r}}} | n_{\text{LL}}, k_y \rangle
= \delta_{k_y', k_y + q_y} \, e^{i q_x l_B^2 (k_y + q_y/2)} F_{n_{\text{LL}}'n_{\text{LL}}}(\boldsymbol{q}),
\end{equation}
where $l_B = \sqrt{\hbar/eB}$ is the magnetic length. The function $F_{n'n}(\boldsymbol{q})$ is the Landau-level form factor, which encodes the overlap between harmonic-oscillator wavefunctions. For $n_{\text{LL}}' \ge n_{\text{LL}}$, it is given by
\begin{eqnarray}
&&F_{n_{\text{LL}}'n_{\text{LL}}}(\boldsymbol{q}) \nonumber\\
&=& \sqrt{\frac{n_{\text{LL}}!}{n_{\text{LL}}'!}} \,
e^{-\frac{q^2 l_B^2}{4}}
\left( \frac{q_x - i q_y}{\sqrt{2}} l_B \right)^{n_{\text{LL}}'-n_{\text{LL}}}
L_{n_{\text{LL}}}^{\,n_{\text{LL}}'-n_{\text{LL}}}\left(\frac{q^2 l_B^2}{2}\right),\nonumber\\
\end{eqnarray}
where $L_{n_{\text{LL}}}^\alpha(x)$ denotes the generalized Laguerre polynomial. The matrix elements for $n_{\text{LL}}' < n_{\text{LL}}$ are obtained from the Hermitian-conjugation relation
\begin{equation}
F_{n_{\text{LL}}'n_{\text{LL}}}(\boldsymbol{q}) = F_{n_{\text{LL}}n_{\text{LL}}'}(-\boldsymbol{q})^*.
\end{equation}

In this Hamiltonian, momentum scattering occurs between $k_y$ values differing by $2\pi/a$. The magnetic Bloch theorem requires $k_y$ to change by $4p\pi/a$ when moving from one magnetic unit-cell boundary to the next. Consequently, upon diagonalization, the original Landau levels are folded and split into $p$ distinct subbands. Strictly speaking, one may count $2p$ subbands, but this doubling arises from an additional folding of the same $p$ subbands without further splitting. This yields the Hofstadter energy spectrum shown in Fig.~\ref{fig:3}(a).

To calculate the optical response, we evaluate interband transitions driven by the local current operator $\hat{\boldsymbol{J}} = -e \hat{\boldsymbol{v}}$. The velocity operators are defined as $\hat{v}_x = \hat{p}_x / m$ and $\hat{v}_y = (\hat{p}_y - eBx) / m$. In a uniform magnetic field, these operators act as ladder operators on the harmonic-oscillator states in the LL basis. Defining the standard annihilation and creation operators as
\begin{equation}
\hat{a} = \frac{1}{\sqrt{2e\hbar B}} \left(\hat{p}_x + i(\hat{p}_y - eBx)\right),
\end{equation}
the velocity operators can be written as
\begin{align}
\hat{v}_x &= v_0 (\hat{a} + \hat{a}^\dagger), \\
\hat{v}_y &= -i v_0 (\hat{a} - \hat{a}^\dagger),
\end{align}
where
\[
\hat{a}^\dagger | n_{\text{LL}}, k_y \rangle
= \sqrt{n_{\text{LL}}+1}\,| n_{\text{LL}}+1, k_y \rangle,
\]
and the characteristic velocity scale is $v_0 = \sqrt{\hbar \omega_c /(2m)}$.

When the periodic potential $V_m$ is introduced, the eigenstates become Bloch states $|u_\alpha(\boldsymbol{k})\rangle$, which are linear combinations of the unperturbed basis states $|n, k_y\rangle$. The velocity matrix elements in the eigenbasis, $\langle u_\alpha(\boldsymbol{k}) | \hat{v}_\mu | u_\beta(\boldsymbol{k}) \rangle$, are calculated directly by expanding the Bloch states in the LL basis and applying the ladder-operator relations above. The interband Berry connection is then obtained from
\begin{equation}
\mathcal{A}^\mu_{\alpha\beta}(\boldsymbol{k}) =
\frac{i \hbar}{\varepsilon_\alpha(\boldsymbol{k}) - \varepsilon_\beta(\boldsymbol{k})}
\langle u_\alpha(\boldsymbol{k}) | \hat{v}_\mu | u_\beta(\boldsymbol{k}) \rangle,
\qquad (\alpha \neq \beta).
\end{equation}
Substituting this Berry connection into the Kubo-Greenwood formula [Eq.~\eqref{eq:band-sigma}], we numerically evaluate the optical Hall conductivity $\tilde{\sigma}_{xy}(\omega)$, the results of which are shown in Fig. 1(d) in the main text.

\section{Magnetic filling $\nu_m = 1$ case of the Hofstadter model}

\begin{figure}[htbp]
\includegraphics[width=0.7\linewidth]{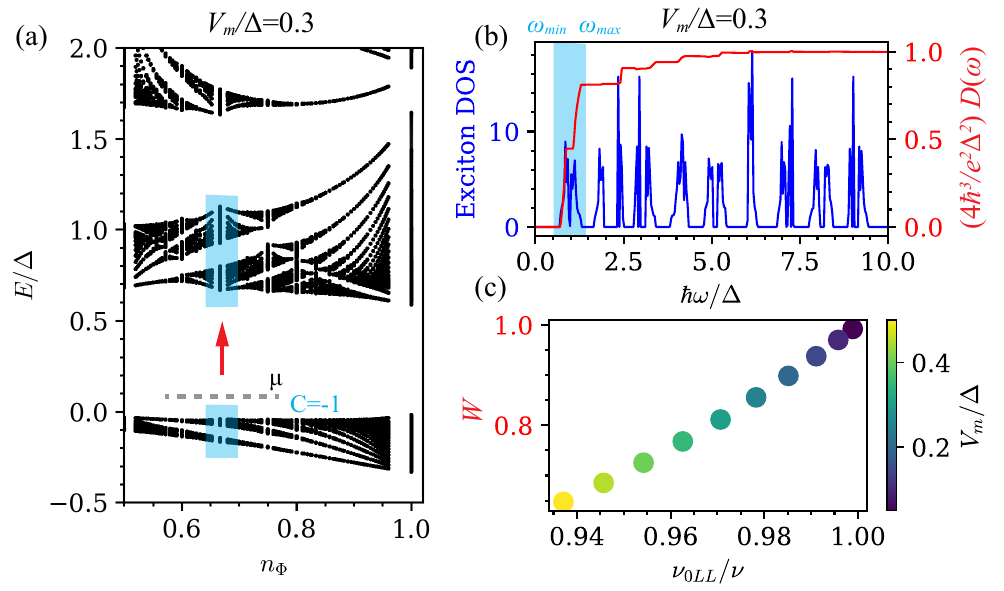}
\caption{\label{fig:3}(a) Hofstadter energy spectrum versus magnetic flux density $n_\Phi = \Phi/\Phi_0$ for a potential strength $V_m/\Delta=0.3$. At $n_\Phi=2/3$, the lowest Landau level (0LL) splits into two subbands carrying a collective Chern number of $C=-1$. The chemical potential $\mu$ (gray dashed line) is placed in the gap above these subbands such that the entire 0LL manifold is fully occupied. $\Delta = \hbar\omega_c$ denotes the pure cyclotron gap at $V_m=0$. The blue shading highlights the specific groups of bands involved in the interband transitions analyzed in (b) and (c). (b) Exciton density of states (DOS) (blue, left axis) and the cumulative first moment normalized by $4\hbar/(e^2 \Delta^2)$ (red, right axis) at $n_\Phi=2/3$. This normalization ensures the integral asymptotically approaches the total electron filling $\nu$ as $\omega \to \infty$. The shaded frequency window $[\omega_{\min}, \omega_{\max}]$ captures the primary transitions between the 0LL and 1LL manifolds. (c) The partial sum rule contribution $W$ (interband weight integrated strictly within the shaded window) plotted against the fractional weight of the 0LL in the occupied ground state, $\nu_{0\mathrm{LL}}/\nu$, as the periodic potential strength $V_m/\Delta$ is varied (color scale).}
\end{figure}

While the main text focuses on the fractional filling of the Hofstadter subbands, in this section, we extend our analysis to the integer quantum Hall regime. Here, the chemical potential lies in the primary gap such that all subbands originating from the lowest Landau level (0LL) are fully occupied (equivalent to a magnetic filling factor of $\nu_m = 1$ for the 0LL manifold). 

The global optical sum rule Eq. (5) in the main text dictates that the total first moment is strictly conserved, depending exclusively on the carrier density and the external magnetic field, independent of the periodic potential:
\begin{equation}
\int_{0}^{\infty} d\omega\,\omega\,\operatorname{Im}\,\tilde{\sigma}_{xy}(\omega) = \frac{\pi n e^3 B}{2m^2} \equiv \frac{e^2 \Delta^2}{4\hbar} \nu,
\label{eq:first_dipole_LL}
\end{equation}
where $\nu = 2\pi \ell_B^2 n$ is the total filling factor and $\Delta = \hbar\omega_c$. As demonstrated in Fig.~\ref{fig:3}(b), the numerically computed cutoff-dependent first moment asymptotically converges to this exact theoretical bound as the integration window is extended to infinity.

To quantify the effect of periodic-potential-induced Landau level (LL) mixing on the low-energy optical response, we define a partial spectral weight $W$. This quantity is integrated over a finite frequency window $[\omega_{\min}, \omega_{\max}]$ encompassing the primary cyclotron transition between the 0LL and 1LL manifolds:
\begin{equation}
W = \frac{4\hbar}{e^2 \Delta^2} \int_{\omega_{\min}}^{\omega_{\max}} d\omega \, \omega \operatorname{Im}\tilde{\sigma}_{xy}(\omega).
\label{eq:w_def}
\end{equation}
In the clean, translationally invariant limit ($V_m \to 0$), inter-LL transitions are strictly confined to the bare cyclotron frequency $\omega_c$, yielding exactly $W = \nu$. However, introducing a finite periodic potential $V_m$ breaks continuous translation symmetry, mixing the 0LL with higher-energy LLs. This mixing transfers spectral weight from the primary cyclotron peak into higher-harmonic transitions residing outside the integration window, causing $W$ to monotonically decrease. 

As shown in Fig.~\ref{fig:3}(c), this reduction in the partial optical spectral weight strongly correlates with the depletion of the ground state's geometric projection onto the ideal 0LL manifold, mathematically defined as $\nu_{0\mathrm{LL}}/\nu = N_e^{-1} \sum_i \langle \Psi | \hat{P}_{0\mathrm{LL}}^{(i)} | \Psi \rangle$. This behavior is qualitatively identical to the fractional subband filling scenario discussed in the main text, thereby confirming that the finite-frequency optical sum rule serves as a robust, filling-independent diagnostic for quantifying the severity of LL mixing in moir\'e and Hofstadter systems.

\section{Correlation between optical weight and band mixing}

In this section, we demonstrate that partial optical weight $W$ could serve as a valid proxy for Landau level (LL) mixing through evaluating the exact algebra of the projected current operators within the primary frequency window.

By the exact sum rule, the total integrated optical Hall weight is proportional to the ground-state expectation value of the current commutator $[\hat{J}_x, \hat{J}_y]$. To simplify the expression, we define \(\hat{J}_\pm = -\frac{1}{2 e v_0}(\hat{J}_x \mp \hat{J}_y)\) such that \([\hat{J}_-, \hat{J}_+] = 1\). The partial weight $W$ measured around the cyclotron frequency captures transitions strictly into the perturbed first Landau level manifold. If $P_{\tilde{1}} = \sum_{f \in \widetilde{\text{1LL}}} |f\rangle\langle f|$ is the exact projector onto this perturbed manifold, the primary weight evaluates to:
\begin{equation}
    W = \langle \tilde{0} | \hat{J}_- P_{\tilde{1}} \hat{J}_+ | \tilde{0} \rangle - \langle \tilde{0} | \hat{J}_+ P_{\tilde{1}} \hat{J}_- | \tilde{0} \rangle,
\end{equation}
This can be further simplified into:
\begin{equation}
    W = \text{Tr}(P_{\tilde{0}} \hat{J}_- P_{\tilde{1}} \hat{J}_+) - \text{Tr}(P_{\tilde{0}} \hat{J}_+ P_{\tilde{1}} \hat{J}_-).
\end{equation}
Here, similarly we define $P_{\tilde{0}}$ as the exact projectors onto the  $\widetilde{\text{0LL}}$ that is occupied. The ground-state projection onto the ideal 0LL is $\text{Tr}(P_{\tilde{0}}P_{0}) = \nu_{0LL}$, and we define $Z = \nu_{0LL} / \nu$ the purity. Without periodic mixing, $Z=1$ and $W = \nu$. 

We decompose the projectors in the unperturbed LL basis into diagonal and off-diagonal components: $P = P^{(d)} + P^{(od)}$. The diagonal components represent the classical statistical weights of each ideal LL within the perturbed manifolds: $P_{\tilde{0}}^{(d)} = \sum_{n_{\text{LL}}} \nu_{n_{\text{LL}}} |n_{\text{LL}}\rangle\langle n_{\text{LL}}|$ (where $\nu_0 = \nu Z$) and $P_{\tilde{1}}^{(d)} = \sum_{n_{\text{LL}}} w_{n_{\text{LL}}} |n_{\text{LL}}\rangle\langle n_{\text{LL}}|$ (where $w_1 \equiv Z_1$). The parameter $\nu_{n_{\text{LL}}}$ explicitly represents the fractional filling of the unperturbed $n_{\text{LL}}$-th LL in the ground state. The off-diagonal components $P^{(od)}$ encapsulate the quantum coherences.

Substituting these decompositions into the trace isolates the classical probability transfer ($d-d$) from the quantum interference ($od-od$).

\textbf{1. Exact Vanishing of Mixed Terms:}
The mixed terms, such as $\text{Tr}(P_{\tilde{0}}^{(d)} \hat{J}_- P_{\tilde{1}}^{(od)} \hat{J}_+)$, vanish identically at all orders of the periodic potential. To prove this, we utilize the cyclic property of the trace to rewrite the term as $\text{Tr}(\hat{J}_+ P_{\tilde{0}}^{(d)} \hat{J}_- P_{\tilde{1}}^{(od)})$. Because the diagonal projector $P_{\tilde{0}}^{(d)}$ is defined purely by states $|n_{\text{LL}}\rangle\langle n_{\text{LL}}|$, and the current operators strictly shift the LL index by one ($\hat{J}_+ |n_{\text{LL}}\rangle \propto |n_{\text{LL}}+1\rangle$ and $\langle n_{\text{LL}}| \hat{J}_- \propto \langle n_{\text{LL}}+1|$), the composite operator $\hat{J}_+ P_{\tilde{0}}^{(d)} \hat{J}_-$ remains strictly diagonal in the unperturbed LL basis.

The trace of any strictly diagonal operator multiplied by a strictly off-diagonal operator ($P_{\tilde{1}}^{(od)}$) is exactly zero. By identical matrix logic, all cross-terms between diagonal and off-diagonal projectors ($(od-d)$ and $(d-od)$) vanish exactly. Consequently, the total partial weight rigorously and cleanly separates into purely classical transfer and purely quantum interference components: $W = W^{(dd)} + W^{(od-od)}$.

\textbf{2. Exact Lowest-Order Diagonal Contribution:}
The diagonal-diagonal trace captures the classical transfer of spectral weight without quantum interference. Using $\hat{J}_+ |n_{\text{LL}}\rangle = \sqrt{n_{\text{LL}}+1} |n_{\text{LL}}+1\rangle$ and $\hat{J}_- |n_{\text{LL}}\rangle = \sqrt{n_{\text{LL}}} |n_{\text{LL}}-1\rangle$, it evaluates exactly to:
\begin{equation}
    W^{(dd)} = \sum_{n_{\text{LL}}=0}^\infty \nu_{n_{\text{LL}}} \left[ (n+1)w_{n+1} - n w_{n-1} \right]
\end{equation}
To evaluate this to the lowest non-trivial order ($\mathcal{O}(V^2)$), we note that $\nu_n \sim \mathcal{O}(V^2)$ for $n \ge 1$, and $w_m \sim \mathcal{O}(V^2)$ for $m \neq 1$. Expanding the sum and discarding terms of $\mathcal{O}(V^4)$ yields:
\begin{equation}
    W^{(dd)} = \nu_0 w_1 - 2\nu_2 w_1 + \mathcal{O}(V^4) = \nu Z Z_1 - 2\nu_2 Z_1
\end{equation}
Because $Z_1 = 1 - \mathcal{O}(V^2)$ and $\nu_2 \sim \mathcal{O}(V^2)$, we can approximate $2\nu_2 Z_1 \approx 2\nu_2$ at this leading order. Furthermore, we can expand the purities $Z = 1 - \delta_0$ and $Z_1 = 1 - \delta_1$, yielding $Z Z_1 \approx Z + Z_1 - 1$. Substituting these simplifies the exact lowest-order classical weight reduction to:
\begin{equation}
    \nu - W^{(dd)} \approx \nu(1-Z) + \nu(1-Z_1) + 2\nu_2
\end{equation}
This mathematically establishes the backbone of the weight reduction: the loss of optical weight is strictly positive and driven simultaneously by the impurities of both participating Landau levels, supplemented by the $2 \to 1$ transition which intrinsically carries a negative sign in the Hall response (contributing the $+2\nu_2$ reduction).

\textbf{3. The Role of Quantum Interference:}
The remaining contribution arises from the pure off-diagonal trace $W^{(od-od)}$. Because both $P_{\tilde{0}}^{(od)}$ and $P_{\tilde{1}}^{(od)}$ are first-order in the periodic potential, this interference term is rigorously $\mathcal{O}(V^2)$. 

Crucially, this means the quantum interference acts at the exact same perturbative order as the classical change in purity $\nu(1-Z)$. Unlike $W^{(dd)}$, the off-diagonal trace lacks a strictly defined sign, depending sensitively on the spatial geometry and phases of the periodic potential's Fourier components. Consequently, while these cross-terms preclude the establishment of a strictly universal numerical bound for all conceivable potentials, they do not qualitatively deviate the scaling. Both the optical weight reduction $\nu - W$ and the ground state impurity $1-Z$ scale cleanly as $\mathcal{O}(V^2)$, mathematically ensuring that the optical weight serves as a robust and monotonically well-behaved proxy for characterizing the extent of Landau level mixing.

\section{Lattice model \label{sec:lattice model}}

In this section, we evaluate the optical Hall sum rule, 
\begin{equation}
    \int_0^\infty d\omega \, \omega \operatorname{Im}\tilde{\sigma}_{xy}(\omega) = -\frac{\pi}{2} \frac{i}{\hbar V} \langle [\hat{J}_x, \hat{J}_y] \rangle
\end{equation}
for a general lattice model. We demonstrate that as long as the kinetic hopping is translationally invariant with a single-site unit cell, and interactions/potentials are strictly position-dependent, the commutator of the current operators vanishes identically.

Consider a generic many-body Hamiltonian on a lattice: $\hat{H} = \hat{K} + \hat{V} + \hat{U}$. 
The potential term $\hat{V} = \sum_i V(\mathbf{r}_i) \hat{n}_i$ and the interaction term $\hat{U} = \sum_{i,j} U(\mathbf{r}_i, \mathbf{r}_j) \hat{n}_i \hat{n}_j$ are completely diagonal in the real-space basis. The uniform current operator is defined via the polarization operator $\hat{\mathbf{P}} = -e \sum_i \mathbf{r}_i \hat{n}_i$ as:
\begin{equation}
    \hat{\mathbf{J}} = -\frac{i}{\hbar} [\hat{H}, \hat{\mathbf{P}}]
\end{equation}
Because $\hat{V}$ and $\hat{U}$ depend only on the local density operators $\hat{n}_i$, they commute exactly with the polarization $\hat{\mathbf{P}}$. Therefore, the current operator is completely independent of the potential and interaction terms; it is entirely determined by the kinetic hopping term:
\begin{equation}
    \hat{\mathbf{J}} = -\frac{i}{\hbar} [\hat{K}, \hat{\mathbf{P}}]
\end{equation}

By assumption, the hopping $\hat{K}$ is translationally invariant with a single-site unit cell. Transforming to momentum space, the kinetic Hamiltonian is strictly diagonal:
\begin{equation}
    \hat{K} = \sum_{\mathbf{k}} \epsilon_{\mathbf{k}} \hat{n}_{\mathbf{k}} = \sum_{\mathbf{k}} \epsilon_{\mathbf{k}} c_{\mathbf{k}}^\dagger c_{\mathbf{k}}
\end{equation}
Evaluating the commutator with the position operator in momentum space ($\hat{\mathbf{P}} = -ie \nabla_{\mathbf{k}}$) yields the standard group-velocity form for the current operators:
\begin{equation}
    \hat{J}_\alpha = -\frac{e}{\hbar} \sum_{\mathbf{k}} \frac{\partial \epsilon_{\mathbf{k}}}{\partial k_\alpha} \hat{n}_{\mathbf{k}}
\end{equation}
for spatial directions $\alpha \in \{x, y\}$. 

Crucially, because $\hat{J}_x$ and $\hat{J}_y$ depend purely on the momentum-space density operators $\hat{n}_{\mathbf{k}}$, and all density operators commute with one another ($[\hat{n}_{\mathbf{k}}, \hat{n}_{\mathbf{k}'}] = 0$), the current operators commute exactly:
\begin{equation}
    [\hat{J}_x, \hat{J}_y] = \frac{e^2}{\hbar^2} \sum_{\mathbf{k}, \mathbf{k}'} \frac{\partial \epsilon_{\mathbf{k}}}{\partial k_x} \frac{\partial \epsilon_{\mathbf{k}'}}{\partial k_y} [\hat{n}_{\mathbf{k}}, \hat{n}_{\mathbf{k}'}] = 0
\end{equation}
This exact operator identity dictates that the global integrated optical Hall weight is strictly zero for such bare lattice models.

\section{Connection to Spontaneous Chern Insulators}

While $\langle [\hat{J}_x, \hat{J}_y] \rangle = 0$ holds strictly for kinetic Hamiltonian with parabolic dispersion or in translation invariant lattice model (see section ~\ref{sec:lattice model}), this does not preclude the emergence of topological phases driven by strong extended interactions, such as a Spontaneous Chern Insulator. In such strongly correlated regimes, the system minimizes interaction energy by spontaneously breaking time-reversal and translation symmetries, dynamically generating chiral loop currents that act as an emergent multi-site magnetic unit cell. 

The apparent paradox of a vanishing exact global Hall sum rule in a Chern insulator, which inherently possesses a non-zero, quantized low-frequency Hall response is cleanly resolved by the separation of energy scales. The finite low-frequency optical Hall weight arising from transitions between the dynamically generated topological bands is exactly canceled by high-frequency Hall transitions across the large interaction-induced gaps (the Mott or charge-density-wave scale). Consequently, the low-energy topological contribution and the high-energy interaction-scale contribution sum identically to zero, perfectly preserving the fundamental bare lattice constraint $\langle [\hat{J}_x, \hat{J}_y] \rangle = 0$.

\end{document}